\documentclass[a4paper,10pt,twoside]{cpc-hepnp}
\usepackage{upgreek,fancyhdr}
\usepackage{multicol}
\usepackage{graphicx}
\usepackage{booktabs}
\usepackage{amssymb,bm,mathrsfs,bbm,amscd}
\usepackage[tbtags]{amsmath}
\usepackage{lastpage}
\usepackage{mathrsfs}
\usepackage{epstopdf}
\usepackage{epsfig}
\usepackage{textcomp}
\usepackage{amsmath}
\allowdisplaybreaks[1]
\usepackage{graphicx}
\usepackage{amssymb}
\usepackage{tikz}
\usepackage{pgfplots}
\usepackage{pgf-umlsd}
\usepackage{slashed}
\begin{document}
\title{Extended Projection Method for Massive Fermion
\thanks{This work was supported by the National Natural Science Foundation of China under Grant No. 11675185.}}
\author{Yefan Wang$^{1,2)}$ \email{wangyefan@ihep.ac.cn}
\quad Zhao Li$^{1,2)}$ \email{zhaoli@ihep.ac.cn}
}
\maketitle
\address{
$^1$ Institute of High Energy Physics, Chinese Academy of Sciences, Beijing 100049, China\\
$^2$ School of Physics Sciences, University of Chinese Academy of Sciences, Beijing 100039, China
}

\begin{abstract}
Tensor reduction is important for multi-loop amplitude calculation.
And the projection method is one of the most popular approaches for tensor reduction.
However, projection method could be problematic for amplitude with massive fermions
due to the inconsistency between helicity and chirality.
We propose an approach to extend the projection method to reduce the loop amplitude containing
fermion chain with two massive spinors.
The extension is achieved by decomposing one of the massive spinors into two specific massless spinors,
''null spinor'' and ''reference spinor''.
Then the extended projection method can be safely implemented for all the processes
including the production of massive fermions.
Finally we present the tensor reduction for the virtual Z boson decaying to top-quark pair to demonstrate our approach.
\end{abstract}



\begin{multicols}{2}

\section{Introduction}

After the discovery of the Higgs boson, the continuous improvement of
experiment accuracy at the CERN Large Hadron Collider (LHC) demands
the precise theoretical predictions to high order corrections.
However, the high order corrections could still be seriously challenging
due to the complicacy of multi-loop Feynman diagrams.
One of the challenging tasks is to reduce the loop amplitude into linear combination of master integrals.

For the one-loop amplitude, a variety of reduction methods has been developed in the past
decades \cite{Passarino:1978jh,Britto:2004nc,delAguila:2004nf,Bern:2005cq,Denner:2005nn,Ossola:2006us,Forde:2007mi,Giele:2008ve,vanHameren:2009vq,Cascioli:2011va}.
The amplitude can be efficiently expressed as linear combinations of one-loop master integrals.
After decades of effort, the one-loop reduction can be carried out with various automated
programs \cite{Ossola:2007ax,Berger:2008sj,vanHameren:2009dr,Hirschi:2011pa,Mastrolia:2010nb,Badger:2012pg,Cullen:2014yla,Peraro:2014cba}.

At the multi-loop level, the achievement of reduction procedure is much harder than the one-loop case.
To improve the efficiency, the reduction for multi-loop amplitude can be conventionally
separated into two steps, i.e. the tensor reduction and the scalar integrals reduction
using integration by part (IBP) identities. For the IBP reduction, many algorithms and
codes have been developed after decades of effort \cite{Laporta:2001dd,Anastasiou:2004vj,Smirnov:2008iw,Smirnov:2014hma,Smirnov:2019qkx,Maierhoefer:2017hyi,Maierhofer:2018gpa,Studerus:2009ye,vonManteuffel:2012np,Lee:2012cn,Lee:2013mka,Georgoudis:2017iza,Bendle:2019csk,Smirnov:2006tz,Lee:2008tj,Schabinger:2011dz,Larsen:2015ped,Boehm:2017wjc,vonManteuffel:2014ixa,Kosower:2018obg}.
On the other hand, tensor reduction is also important.
During past decades, many algorithms for tensor reduction have been
proposed \cite{Binoth:2002xg,Glover:2004si,Tarasov:1996br,Mastrolia:2011pr,Badger:2012dp,Zhang:2012ce,Liu:2018dmc,Wang:2019mnn}.
For some complicated processes, such as full next-to-next-to-leading order QCD correction to single-top
production \cite{Assadsolimani:2014oga}, the increasing number of form factors makes
the coefficients hard to obtain. And for two-loop five-gluon or six-gluon amplitude,
the corresponding system of equations for the coefficients can be very complicated
to be solved \cite{Peraro:2019cjj,Chen:2019wyb}.
Moreover, some complicated processes, e.g. $e^{+}e^{-}\rightarrow Z^{\star}\rightarrow t\bar{t}$,
can confront serious problem during tensor reduction. This indicates that the further investigation on tensor reduction is still needed.

The projection method \cite{Binoth:2002xg,Glover:2004si} is one of the most popular approaches for tensor reduction.
During past decades, many important researches have been done by using projection method, such as high order
QCD corrections to the Higgs production \cite{Gehrmann:2011aa,Borowka:2016ehy,Melnikov:2017pgf,Jones:2018hbb}
and vector boson production \cite{Gehrmann:2011ab,Gehrmann:2015ora}.
However, for some processes containing fermion chain with two massive spinors, the projection method could be sabotaged by
the inconsistency between helicity and chirality, which will be explicitly shown in next section.

In this paper, based on the massive spinor decomposition \cite{Kosower:2004yz,Schwinn:2005pi,Feng:2011np},
we propose to extend the projection method to reduce the loop amplitude for any process including production of massive fermions. The massive spinor can be decomposed by defining ''null spinor'' and ''reference spinor'',
which have completely different formulas of equation of motion and polarization summation compared to the regular spinor.
Then for the processes containing massive spinors the projection method can be safely used.

This paper is organized as follows.
In section 2 we briefly review the standard projection method
and demonstrate the problem due to the lack of massless spinor.
In section 3 based on the massive spinor decomposition we introduce ''null spinor'' and ''reference spinor'',
which can be used to extend the projection method for all processes.
In section 4 and 5, we take one-loop and two-loop diagrams for virtual Z boson decaying to top-quark pair to demonstrate the effectiveness of our approach, respectively. The conclusion is presented in the last section.

\section{Standard Projection method}

In projection method, the loop amplitude can be expressed as the linear combination of several monomials.
In each monomial the Lorentz structure is composed of
the spinors and polarization vectors associated with the contracted momenta,
while the remnant factors including coupling constants and scalar products of momenta will not affect the tensor reduction.
Then by trimming off the chirality for spinors from the Lorentz structure we can obtain the primitive amplitude.
Therefore, the loop amplitude can be decomposed as
\begin{eqnarray}
\mathcal{A}=\sum_{p,X} C_{p,X} \mathcal{M}_{p,X},
\end{eqnarray}
where $X$ is chirality index and $p$ indicates the different primitive amplitudes.
$\mathcal{M}_{p,X}$ is the Lorentz structure for certain chirality $X$,
and $C_{p,X}$ is the relevant coefficient.

For convenience we can define a map for each primitive amplitude from helicity state $H$ to chirality state $X$
\begin{equation}
f_p: H \mapsto X, {\mathrm~ s.t.~} \mathbb{P}_H \mathcal{M}_{p,X} \neq 0,
\end{equation}
where $\mathbb{P}_H$ is the helicity projection operator. For instance, the explicit map for the primitive amplitude
\begin{equation}
\bar u(k_1) \slashed q v(k_2),
\end{equation}
where $k_1^2=k_2^2=0$, can written as
\begin{equation}
f_p(+-) = LR, \quad f_p(-+) = RL.
\end{equation}
Furthermore, for the primitive amplitude with one massive spinor, for example
\begin{equation}
\bar u(k_1) \slashed q v(k_2),
\end{equation}
where $k_1^2\neq0$ and $k_2^2=0$, the map still can be constructed as
\begin{equation}
f_p(--) = f_p(+-) = LR, \quad f_p(++) = f_p(-+) = RL.
\end{equation}
Obviously, the map can be established based on at least one massless spinor, which has equivalence relation between helicity and chirality and further can be used to fix the chirality for the relevant fermion chain by using anti-commute $\gamma_5$ scheme.
However, the map does not exist for the primitive amplitude containing two massive spinors in the same fermion chain,
which is just the case that projection method fails.

Therefore we can obtain
\begin{align}
\mathcal{M}_{p,X} = \sum_H \mathbb{P}_H \mathcal{M}_{p,X}
= \sum_{H=f_p^{-1}(X)} \delta_{X,f_p(H)}\mathcal{M}_{p,H},
\end{align}
where $H$ indicates certain helicity state for the spinors.
$f_p^{-1}(X)$ is not the inverse of $f_p$ but only represents the set of helicity states that can be mapped to certain chirality state $X$.

Consequently the amplitude $\mathcal{A}$ can be expressed as linear combinations of helicity primitive amplitudes,
\begin{align}
\mathcal{A} =\sum_{p,H}C_{p,X=f(H)} ( \mathbb{P}_H \mathcal{M}_{p} ).
\end{align}
Now one needs only the tensor reduction on the primitive amplitude $\mathcal{M}_p$,
and the loop amplitude $\mathcal{A}$ can be reconstructed by implementing the helicity projection $\mathbb{P}_H$
and summing up all primitive amplitude choices $p$ and helicity states $H$.
Meanwhile the above derivation also presents the formula for helicity amplitude
\begin{align}
\mathcal{A}_H =\sum_{p}C_{p,X=f(H)} ( \mathbb{P}_H \mathcal{M}_{p} ).
\end{align}
In order to make tensor reduction on the primitive amplitude $\mathcal{M}_p$,
one needs to find a complete set of linear independent form factors $\{F_i\}$.
Then $\mathcal{M}_p$ can be projected to form factors
\begin{align}
\mathcal{M}_p =\sum_{i} d_{i,p} F_i.
\label{primitiveTR}
\end{align}
To obtain the explicit expression of coefficient $d_{i,p}$,
both sides of Eq.(\ref{primitiveTR}) can be multiplied by
the conjugate form factor $F^\dagger_j$.
Then the form factor matrix can be defined as
\begin{align}
M_{ij}\equiv F_iF_j^\dagger.
\end{align}
The coefficient $d_{i,p}$ can be obtained from the inversion of matrix $M$
\begin{align}
d_{i,p} = \sum_{j} (M^{-1})_{ij}\mathcal{M}_p F_j^\dagger.
\end{align}
Finally the amplitude $\mathcal{A}$ can be expressed by the linear combinations of $\mathbb{P}_{H} F_i$
\begin{equation}
\mathcal{A}= \sum_{i,H} c_{i,H}(\mathbb{P}_{H} F_i),
\end{equation}
where
\begin{equation}
c_{i,H} = \sum_{p}d_{i,p}C_{p,X=f(H)}.
\end{equation}
Here $\mathbb{P}_{H} F_i$ is independent of loop momenta, and it can be further expressed in spinor representation. And its coefficient $c_{i,H}$ contains the scalar integrals, which can be further reduced by IBP method.

\section{Projection method for massive fermion chain}

In the above section it can be seen that the problem in projection method is due to the lack of massless spinor.
One of the convenient approaches is
to decompose the massive momentum $k$ by introducing reference momentum $k_r$ \cite{Kosower:2004yz,Schwinn:2005pi,Feng:2011np},
\begin{eqnarray}
k = k_0+\dfrac{m^2_k}{2 k_{0}\cdot k_{r}}k_{r},
\end{eqnarray}
where $k^2_{0} = k^2_{r} = 0$ and $k^2 = m^2_k$. And the massive spinor can be decomposed as
\begin{align}
u^+(k,m_k)&=\left|k_{0} \right \rangle+\dfrac{m_k}{\left[k_{0}k_{r}\right]}\left|k_{r}\right],\nonumber\\
u^-(k,m_k)&=\left|k_{0} \right ]+\dfrac{m_k}{\left\langle k_{0}k_{r}\right\rangle}\left|k_{r}\right\rangle,\nonumber\\
v^+(k,m_k)&=\left|k_{0} \right ]-\dfrac{m_k}{\left\langle k_{0}k_{r}\right\rangle}\left|k_{r}\right\rangle,\nonumber\\
v^-(k,m_k)&=\left|k_{0} \right \rangle-\dfrac{m_k}{\left[k_{0}k_{r}\right]}\left|k_{r}\right].
\end{align}
Here we found in fact that on the right-hand side of each above equation the two terms can be defined as two special spinors so that
\begin{align}
u(k,m_k)&=u_{0}(k_0)+ u_{r}(k_{r}),\nonumber\\
v(k,m_k)&=v_{0}(k_0)+ v_{r}(k_{r}).
\end{align}
Explicitly
\begin{align}
u^+_{0}(k_0) &\equiv \left|k_{0} \right \rangle,\quad u^+_{r}(k_{r}) \equiv \dfrac{m_k}{\left[k_{0}k_{r}\right]}\left|k_{r}\right],\nonumber\\
u^-_{0}(k_0) &\equiv \left|k_{0} \right ],\quad u^-_{r}(k_{r}) \equiv \dfrac{m_k}{\left\langle k_{0}k_{r}\right\rangle}\left|k_{r}\right\rangle,\nonumber\\
v^+_{0}(k_0) &\equiv \left|k_{0} \right ],\quad v^+_{r}(k_{r}) \equiv -\dfrac{m_k}{\left\langle k_{0}k_{r}\right\rangle}\left|k_{r}\right\rangle,\nonumber\\
v^-_{0}(k_0) &\equiv \left|k_{0} \right \rangle,\quad v^-_{r}(k_{r}) \equiv -\dfrac{m_k}{\left[k_{0}k_{r}\right]}\left|k_{r}\right].
\end{align}
Now we define $u_r$ and $v_r$ as "reference spinors", although they are not orthogonal to the "null spinors" $u_0$ and $v_0$.
The polarization summation formula can be written as
\begin{align}
u^+_{0}(k_0) \bar{u}^+_{0}(k_0) + u^-_{0}(k_0) \bar{u}^-_{0}(k_0) &= \slashed{k_0},\nonumber\\
v^+_{0}(k_0) \bar{v}^+_{0}(k_0) + v^-_{0}(k_0) \bar{v}^-_{0}(k_0) &= \slashed{k_0},\nonumber\\
u^+_{r}(k_r) \bar{u}^+_{r}(k_r) + u^-_{r}(k_r) \bar{u}^-_{r}(k_r) &= \dfrac{m^2_k}{2 k_0 \cdot k_r} \slashed{k_r},\nonumber\\
v^+_{r}(k_r) \bar{v}^+_{r}(k_r) + v^-_{r}(k_r) \bar{v}^-_{r}(k_r) &= \dfrac{m^2_k}{2 k_0 \cdot k_r} \slashed{k_r},\nonumber\\
u^+_{0}(k_0) \bar{u}^+_{r}(k_r) + u^-_{0}(k_0) \bar{u}^-_{r}(k_r) &= \dfrac{1}{m_k}\slashed{k_0}\slashed{k},\nonumber\\
u^+_{r}(k_r) \bar{u}^+_{0}(k_0) + u^-_{r}(k_r) \bar{u}^-_{0}(k_0) &= \dfrac{1}{m_k}\slashed{k}\slashed{k_0},\nonumber\\
v^+_{0}(k_0) \bar{v}^+_{r}(k_r) + v^-_{0}(k_0) \bar{v}^-_{r}(k_r) &= -\dfrac{1}{m_k}\slashed{k_0}\slashed{k},\nonumber\\
v^+_{r}(k_r) \bar{v}^+_{0}(k_0) + v^-_{r}(k_r) \bar{v}^-_{0}(k_0) &= -\dfrac{1}{m_k}\slashed{k}\slashed{k_0}.
\end{align}
Besides, a set of non-trivial Dirac equations for null spinors and reference spinors can be found as
\begin{align}
\slashed{k}u_{0}(k_0) &= m_k\, u_{r}(k_r),\nonumber\\
\slashed{k}u_{r}(k_r) &= m_k\, u_{0}(k_0),\nonumber\\
\slashed{k}v_{0}(k_0) &= -m_k\, v_{r}(k_r),\nonumber\\
\slashed{k}v_{r}(k_r) &= -m_k\, v_{0}(k_0).
\end{align}

Then since one of the fermion spinors becomes massless,
the map from helicity state to chirality state for primitive amplitude can be constructed.
Finally the projection method can be directly implemented on the decomposed primitive amplitudes.

\section{One-loop Example}

In this section, we take a one-loop amplitude reduction to demonstrate our method. We consider one-loop diagram for $Z^* (k_1)\rightarrow t(k_2)\bar{t}(k_3)$ shown in Figure 1. Here we define $Q^2 \equiv k_1^2$ and $m_t$ as the mass of top-quark. The diagram is plotted by using Jaxodraw \cite{Binosi:2003yf} based on Axodraw \cite{Vermaseren:1994je}.
\begin{center}
	\includegraphics[width=6cm,height=4cm]{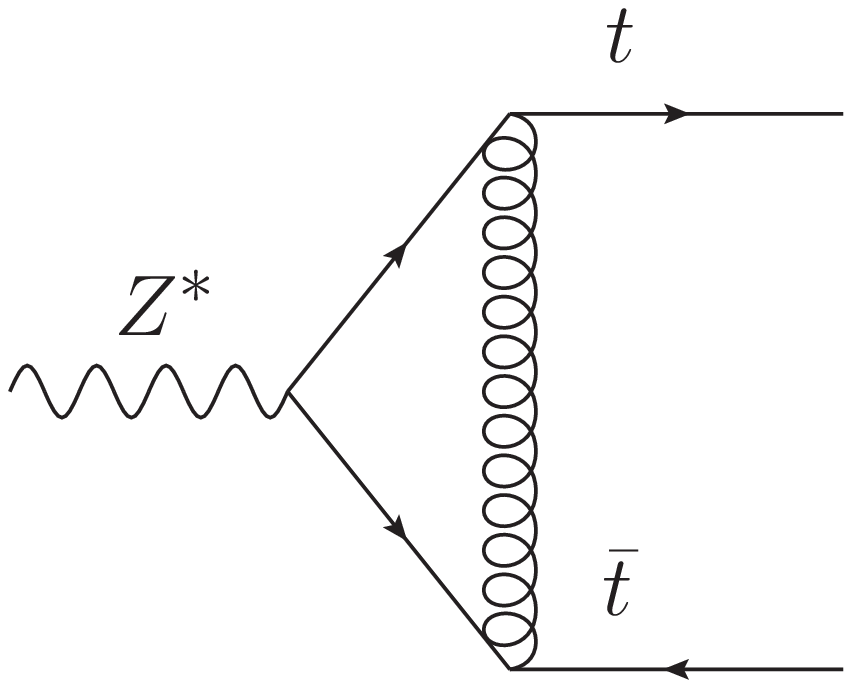}
	\figcaption{\label{one-loop} One-loop triangle diagram for virtual Z boson decaying to top-quark pair.}
\end{center}
Its relevant amplitude can be written as
\begin{equation}
{\mathcal A}= \int \mathrm{d}^D q
\dfrac{N(q,k_1,k_2,k_3)}{
	{\mathcal D}_1
	{\mathcal D}_2
	{\mathcal D}_3
},
\end{equation}
where $q$ is loop momentum. $N(q,k_1,k_2,k_3)$ is the numerator of amplitude. And the denominators are
\begin{align}
{\mathcal D}_1 &= q^2 -m_t^2,\nonumber\\
{\mathcal D}_2 &= (q-k_1)^2 -m_t^2,\nonumber\\
{\mathcal D}_3 &= (q-k_3)^2.
\end{align}
Without loss of generality, here we only consider the right handed current of $Z^*$. Then the numerator can be chosen as
\begin{align}
N_R(q,k_1,k_2,k_3) =&
-g^2_s \, \bar{u}(k_2,m_t)\gamma^a(\slashed{k_1}-\slashed{q}+m_t)\nonumber\\&\times\slashed{\varepsilon}P_R(\slashed{q}+m_t)\gamma^a{v}(k_3,m_t).
\end{align}
Following the approach described in section 2, we choose to decompose $v(k_3,m_t)$,
\begin{align}
{v}({k_3,m_t}) &= {v}_0(k_{30})+{v}_r(k_{3r}),\nonumber\\
k_3&=k_{30}+\dfrac{m^2_t}{2k_{30}\cdot k_{3r}}k_{3r}.
\end{align}
Since the reference momentum $k_{3r}$ has been added, we define two scalar products,
\begin{align}
s_1 &\equiv k_{3r}\cdot k_{30}, \nonumber\\
s_2 &\equiv k_{3r}\cdot k_{1}.
\end{align}
Based on the monomials in the original amplitude, we obtain 12 primitive amplitudes,
\begin{align}
\mathcal{M}_1 &= \int \mathrm{d}^D q
\dfrac{\bar{u}(k_2,m_t)\slashed{\varepsilon} v_{0}(k_{30})}{
	{\mathcal D}_1
	{\mathcal D}_2
	{\mathcal D}_3
},\nonumber\\
\mathcal{M}_{2} &=\int \mathrm{d}^D q
\dfrac{\bar{u}(k_2,m_t)\slashed{\varepsilon} v_{r}(k_{3r})}{
	{\mathcal D}_1
	{\mathcal D}_2
	{\mathcal D}_3
},\nonumber\\
\mathcal{M}_{3} &=\int \mathrm{d}^D q
\dfrac{\bar{u}(k_2,m_t)\slashed{\varepsilon}\slashed{q}v_{0}(k_{30})}{
	{\mathcal D}_1
	{\mathcal D}_2
	{\mathcal D}_3
},\nonumber\\
\mathcal{M}_{4} &=\int \mathrm{d}^D q
\dfrac{\bar{u}(k_2,m_t)\slashed{\varepsilon}\slashed{q}v_{r}(k_{3r})}{
	{\mathcal D}_1
	{\mathcal D}_2
	{\mathcal D}_3
},\nonumber\\
\mathcal{M}_{5}&=\int \mathrm{d}^D q
\dfrac{(k_3\cdot\varepsilon)\bar{u}(k_2,m_t)v_{0}(k_{30})}{
	{\mathcal D}_1
	{\mathcal D}_2
	{\mathcal D}_3
},\nonumber\\
\mathcal{M}_{6}&=\int \mathrm{d}^D q
\dfrac{(k_3\cdot\varepsilon)\bar{u}(k_2,m_t)v_{r}(k_{3r})}{
	{\mathcal D}_1
	{\mathcal D}_2
	{\mathcal D}_3
},\nonumber\\
\mathcal{M}_{7}&=\int \mathrm{d}^D q
\dfrac{(k_3\cdot\varepsilon)\bar{u}(k_2,m_t)\slashed{q}v_{0}(k_{30})}{
	{\mathcal D}_1
	{\mathcal D}_2
	{\mathcal D}_3
},\nonumber\\
\mathcal{M}_{8}&=\int \mathrm{d}^D q
\dfrac{(k_3\cdot\varepsilon)\bar{u}(k_2,m_t)\slashed{q}v_{r}(k_{3r})}{
	{\mathcal D}_1
	{\mathcal D}_2
	{\mathcal D}_3
},\nonumber\\
\mathcal{M}_{9}&=\int \mathrm{d}^D q
\dfrac{(q\cdot\varepsilon)\bar{u}(k_2,m_t)v_{0}(k_{30})}{
	{\mathcal D}_1
	{\mathcal D}_2
	{\mathcal D}_3
},\nonumber\\
\mathcal{M}_{10}&=\int \mathrm{d}^D q
\dfrac{(q\cdot\varepsilon)\bar{u}(k_2,m_t)v_{r}(k_{3r})}{
	{\mathcal D}_1
	{\mathcal D}_2
	{\mathcal D}_3
},\nonumber\\
\mathcal{M}_{11}&=\int \mathrm{d}^D q
\dfrac{(q\cdot\varepsilon)\bar{u}(k_2,m_t)\slashed{q}v_{0}(k_{30})}{
	{\mathcal D}_1
	{\mathcal D}_2
	{\mathcal D}_3
},\nonumber\\
\mathcal{M}_{12}&=\int \mathrm{d}^D q
\dfrac{(q\cdot\varepsilon)\bar{u}(k_2,m_t)\slashed{q}v_{r}(k_{3r})}{
	{\mathcal D}_1
	{\mathcal D}_2
	{\mathcal D}_3
}.
\end{align}
According to the Lorentz structures, it can be found that one needs four linear independent form factors,
\begin{align}
F_{1} &= \bar{u}(k_2,m_t)\slashed{\varepsilon} v_{0}(k_{30}),\nonumber\\
F_{2} &= \bar{u}(k_2,m_t)\slashed{\varepsilon} v_{r}(k_{3r}),\nonumber\\
F_{3} &=(k_3\cdot\varepsilon)\bar{u}(k_2,m_t)v_{0}(k_{30}),\nonumber\\
F_{4} &=(k_3\cdot\varepsilon)\bar{u}(k_2,m_t)v_{r}(k_{3r}).
\end{align}
Then primitive amplitudes can be projected to form factors
\begin{align}
\mathcal{M}_{p} = d_{1,p}F_{1}+d_{2,p}F_{2}+d_{3,p}F_{3}+d_{4,p}F_{4}.
\end{align}
Since the spinors $v_{0}(k_{30})$ and $v_{r}(k_{3r})$ are not completely independent, some relations can be obtained to cancel reference momenta in final result. For example we note that the only difference between $\mathcal{M}_{3}$ and $\mathcal{M}_{4}$ is the last spinor,
\begin{align}
\mathcal{M}_{3} &=\bar{u}(k_2,m_t)\slashed{\varepsilon}\slashed{q}v_{0}(k_{30}),\nonumber\\
\mathcal{M}_{4} &=\bar{u}(k_2,m_t)\slashed{\varepsilon}\slashed{q}v_{r}(k_{3r}).
\end{align}
By observing the symmetry between $v_{0}(k_{30})$ and $v_{r}(k_{3r})$
\begin{align}
\slashed{k_3}v_{0}(k_{30}) &= -m_t v_{r}(k_{3r}),\nonumber\\
\slashed{k_3}v_{r}(k_{3r}) &= -m_t v_{0}(k_{30}),
\end{align}
we can find the relations
\begin{eqnarray}
d_{3,1} = d_{4,2},\quad
d_{3,2} = d_{4,1},\quad
d_{3,3} = d_{4,4},\quad
d_{3,4} = d_{4,3}.
\end{eqnarray}
By using these relations we can obtain
\begin{align}
{k_{3r}} \cdot {{q}} =&\dfrac{2}{Q^4-4 {m^2_t} {Q^2}}
\{({k_2}\cdot {{q}}  ) ({Q^2} {s_1}-2 {m^2_t} {s_2})\nonumber\\
&+({k_3}\cdot {{q}}  )({Q^2} ({s_2}-{s_1})-2 {m^2_t} {s_2})\}.
\end{align}
After the projection on primitive amplitudes, in each form factor we also need to decompose $k_2$ as
\begin{align}
{u}({k_2,m_t}) &= {u}_0(k_{20})+{u}_r(k_{2r}),\nonumber\\
k_2&=k_{20}+\dfrac{m^2_t}{2k_{20}\cdot k_{2r}}k_{2r}.
\end{align}
Then the explicit expressions for $\mathbb{P}_H F_i$ can be written in spinor representation as
\begin{align}
\mathbb{P}_{-+} F_1 &= \left\langle k_{20}|\slashed{\varepsilon}|k_{30}\right],\nonumber\\
\mathbb{P}_{++} F_1 &= \dfrac{m_t}{\left\langle k_{2r}|k_{20}\right\rangle}\left\langle k_{2r}|\slashed{\varepsilon}|k_{30}\right],\nonumber\\
\mathbb{P}_{++} F_2 &= -\dfrac{m_t}{\left\langle k_{30}|k_{3r}\right\rangle} \left[k_{20}|\slashed{\varepsilon}|k_{3r}\right\rangle,\nonumber\\
\mathbb{P}_{-+} F_2 &= -\dfrac{m_t^2}{\left[k_{2r}|k_{20}\right]\left\langle k_{30}|k_{3r}\right\rangle} \left[k_{2r}|\slashed{\varepsilon}|k_{3r}\right\rangle,\nonumber\\
\mathbb{P}_{++} F_3 &= (k_3\cdot\varepsilon)\left[k_{20}|k_{30}\right],\nonumber\\
\mathbb{P}_{-+} F_3 &= (k_3\cdot\varepsilon)\dfrac{m_t}{\left[k_{2r}|k_{20}\right]}\left[k_{2r}|k_{30}\right],\nonumber\\
\mathbb{P}_{-+} F_4 &= -(k_3\cdot\varepsilon) \dfrac{m_t}{\left\langle k_{30}|k_{3r}\right\rangle} \left\langle k_{20}|k_{3r}\right\rangle,\nonumber\\
\mathbb{P}_{++} F_4 &= -(k_3\cdot\varepsilon) \dfrac{m_t^2}{\left\langle k_{2r}|k_{20}\right\rangle\left\langle k_{30}|k_{3r}\right\rangle}\left\langle k_{2r}|k_{3r}\right\rangle,\nonumber\\
\mathbb{P}_{+-} F_1 &= \left[k_{20}|\slashed{\varepsilon}| k_{30} \right\rangle ,\nonumber\\
\mathbb{P}_{--} F_1 &= \dfrac{m_t}{\left[ k_{2r}|k_{20}\right]} \left[k_{2r}|\slashed{\varepsilon} |k_{30}\right\rangle,\nonumber\\
\mathbb{P}_{--} F_2 &= -\dfrac{m_t}{\left[ k_{30}|k_{3r}\right]} \left\langle k_{20}|\slashed{\varepsilon}|k_{3r}\right],\nonumber\\
\mathbb{P}_{+-} F_2 &= -\dfrac{m_t^2}{\left\langle k_{2r}|k_{20}\right\rangle \left[ k_{30}|k_{3r}\right]} \left\langle k_{2r}|\slashed{\varepsilon}| k_{3r}\right],\nonumber\\
\mathbb{P}_{--} F_3 &= (k_3\cdot\varepsilon)\left\langle k_{20}|k_{30} \right\rangle,\nonumber\\
\mathbb{P}_{+-} F_3 &= (k_3\cdot\varepsilon)\dfrac{m_t}{\left\langle k_{2r}|k_{20}\right\rangle} \left\langle k_{2r}|k_{30} \right\rangle,\nonumber\\
\mathbb{P}_{+-} F_4 &= -(k_3\cdot\varepsilon) \dfrac{m_t}{\left[ k_{30}|k_{3r}\right]} \left[ k_{20}|k_{3r}\right],\nonumber\\
\mathbb{P}_{--} F_4 &= -(k_3\cdot\varepsilon) \dfrac{m_t^2}{\left[ k_{2r}|k_{20}\right]\left[ k_{30}|k_{3r}\right]} \left[ k_{2r}|k_{3r}\right].
\end{align}
The coefficients $c_{i,H}$ are
\end{multicols}
\ruleup
\begin{align}
c_{1,-+}  =&c_{1,++}= c_{2,--} =c_{2,+-} \nonumber\\=  &\dfrac{-g^2_s}{Q^4-4  m^2_{t} Q^2} \int \mathrm{d}^D q
\dfrac{1}{
	{\mathcal D}_1
	{\mathcal D}_2
	{\mathcal D}_3
}\big\{8 m^4_{t} Q^2-2 m^2_{t} Q^4+\left(8 (D-4) m^4_{t}-4 m^2_{t} Q^2\right) (k_2 \cdot q )+(8 (D-4) m^4_{t}-4 (D-5) m^2_{t} Q^2)(k_3 \cdot q )\big\},\nonumber\\
\nonumber\\
c_{2,++}  =&c_{2,-+}= c_{1,+-} =c_{1,--} \nonumber\\=   & \dfrac{-g^2_s}{Q^4-4  m^2_{t} Q^2} \int \mathrm{d}^D q
\dfrac{1}{
	{\mathcal D}_1
	{\mathcal D}_2
	{\mathcal D}_3
}\big\{(D-4) m^2_{t} Q^2 \left(Q^2-4 m^2_{t}\right)+(8 (D-4) m^4_{t}-4 (D-1)m^2_{t} Q^2+4 Q^4)(k_2 \cdot q )\nonumber\\&+\left(8 (D-4) m^4_{t}+4 (7-2 D) m^2_{t} Q^2+2 (D-4) Q^4\right) (k_3 \cdot q )+\left(16 m^2_{t}-8 Q^2\right)(k_2 \cdot q ) (k_3 \cdot q )\nonumber\\&+8 m^2_{t} (k_2 \cdot q )^2+8 m^2_{t} (k_3 \cdot q )^2
-(D-4) Q^2\left(Q^2-4 m^2_{t}\right) q^2\big\},\nonumber\\
\nonumber\\
c_{3,++}  =& c_{3,-+} = c_{4,+-} = c_{4,--} \nonumber\\=   & \dfrac{-g^2_s}{\left(Q^3-4 m^2_t Q\right)^2}\int \mathrm{d}^D q
\dfrac{1}{
	{\mathcal D}_1
	{\mathcal D}_2
	{\mathcal D}_3
}\big\{2(D-4) m_{t} \left(Q^3-4 m^2_{t} Q\right)^2+(-32 (D-4) m_{t}^5+8 \,D\, m^3_{t} Q^2-8 m_{t} Q^4)(k_2 \cdot q )\nonumber\\&-4 m_{t} \left(4 m^2_{t}-Q^2\right) \left(2 (D-4) m^2_{t}-(D-6) Q^2\right) (k_3 \cdot q )+\left(8\, D \,m_{t} Q^2-32 m^3_{t}\right)(k_2 \cdot q ) (k_3 \cdot q )\nonumber\\&+\left(16 (D-3) m^3_{t}-8 (D-2) m_{t} Q^2\right) (k_2 \cdot q )^2-16 (D-1) m^3_{t} (k_3 \cdot q)^2-4 m_{t} Q^2\left(Q^2-4 m^2_{t}\right) q^2\big\},\nonumber\\
\nonumber
\\
c_{4,-+}  =& c_{4,++} = c_{3,--} = c_{3,+-} \nonumber\\=   & \dfrac{-g^2_s}{\left(Q^3-4 m^2_t Q\right)^2}\int \mathrm{d}^D q
\dfrac{1}{
	{\mathcal D}_1
	{\mathcal D}_2
	{\mathcal D}_3
}\big\{-8 (D-4) m^3_{t} \left(4 m^2_{t}-Q^2\right) (k_2 \cdot q )-4 (D-4) m_{t}  (8 m^4_{t}-6 m^2_{t} Q^2
+Q^4)
(k_3 \cdot q )\nonumber\\&+\left(8\,D\,m_{t} Q^2-32 m^3_{t}\right) (k_2 \cdot q ) (k_3 \cdot q )-16 (D-1) m^3_{t}(k_2 \cdot q )^2
+\left(16 (D-3) m^3_{t}-8 (D-2) m_{t} Q^2\right) (k_3 \cdot q )^2\nonumber\\&-4 m_{t} Q^2  \left(Q^2
-4 m^2_{t}\right)q^2\big\}.
\end{align}
\ruledown
\vspace{0.5cm}
\begin{multicols}{2}
The above result has been cross checked by Tarasov's method \cite{Tarasov:1996br} using FaRe \cite{Fiorentin:2015vha} and LiteRed \cite{Lee:2013mka}.

\section{Two-loop Example }

\begin{center}
	\includegraphics[width=6cm,height=4cm]{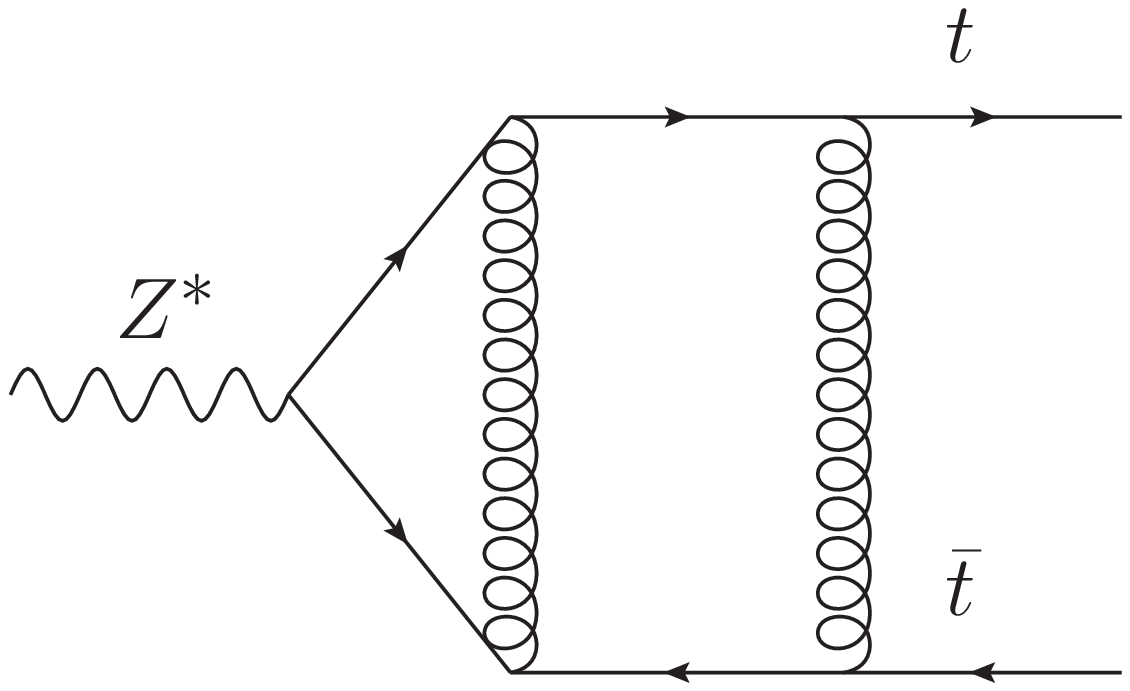}
	\figcaption{\label{two-loop} Two-loop triangle diagram for virtual Z boson decaying to top-quark pair.}
\end{center}

In this section we take one two-loop diagram to demonstrate that our method can be used in the higher order correction. One typical two-loop diagram for $Z^*(k_1)\rightarrow t(k_2)\bar{t}(k_3)$ is shown in Figure 2. Its relevant amplitude can be written as
\begin{equation}
{\mathcal A}= \int \mathrm{d}^D q_1 \mathrm{d}^D q_2
\dfrac{N(q_1,q_2,k_1,k_2,k_3)}{
	{\mathcal D}_1
	{\mathcal D}_2
	{\mathcal D}_3
	{\mathcal D}_4
	{\mathcal D}_5
	{\mathcal D}_6
},
\end{equation}
where $q_1$ and $q_2$ are the loop momenta. The $N(q_1,q_2,k_1,k_2,k_3)$ is the numerator of the two-loop amplitude. The loop denominators are
\begin{align}
{\mathcal D}_1 &= (q_2+k_2)^2 -m_t^2,\nonumber\\
{\mathcal D}_2 &= (q_1)^2 - m_t^2,\nonumber\\
{\mathcal D}_3 &= (q_1-k_1)^2 ,\nonumber\\
{\mathcal D}_4 &= (q_2-k_3)^2 - m_t^2,\nonumber\\
{\mathcal D}_5 &= (q_2)^2,\nonumber\\
{\mathcal D}_6 &= (q_1-q_2-k_2)^2.
\end{align}
And in complete integral family for the above two-loop amplitude, one additional denominator is needed
\begin{align}
{\mathcal D}_7 = (q_1-k_2)^2.
\end{align}
Without loss of generality, here we only consider the right handed current of $Z^*$. Then the numerator can be chosen as
\begin{align}
N_R(&q_1,q_2,k_1,k_2,k_3) =
g^4_s \bar{u}(k_2)\gamma^a(\slashed{q_2}+\slashed{k_2}+m_t)\gamma^b(\slashed{q_1}+m_t)\nonumber\\
&\times \slashed{\varepsilon}P_R(\slashed{k_1}-\slashed{q_1}+m_t)
\gamma^b (\slashed{k_3}-\slashed{q_2}+m_t)\gamma^a v(k_3).
\end{align}
We choose to decompose $v(k_3,m_t)$ and define two additional scalar products,
\begin{align}
s_1 &\equiv k_{3r}\cdot k_{30},\nonumber\\
s_2 &\equiv k_{3r}\cdot k_{1} .
\end{align}
Then we can obtain four linear independent form factors
\begin{align}
F_{1} &= \bar{u}(k_2,m_t)\slashed{\varepsilon} v_{0}(k_{30}),\nonumber\\
F_{2} &= \bar{u}(k_2,m_t)\slashed{\varepsilon} v_{r}(k_{3r}),\nonumber\\
F_{3} &=(k_3\cdot\varepsilon)\bar{u}(k_2,m_t)v_{0}(k_{30}),\nonumber\\
F_{4} &=(k_3\cdot\varepsilon)\bar{u}(k_2,m_t)v_{r}(k_{3r}).
\end{align}
Similarly to the one-loop case in last section, we can find relations,
\begin{align}
{k_{3r}} \cdot {{q_1}} =& \dfrac{2}{Q^4-4 {m^2_t} Q^2}
\{({k_2}\cdot {{q_1}}  ) ({Q^2} {s_1}-2 {m^2_t} {s_2})\nonumber\\
&+({k_3}\cdot {{q_1}}  )({Q^2} ({s_2}-{s_1})-2 {m^2_t} {s_2})\},\\
\nonumber\\
{k_{3r}} \cdot {{q_2}} =& \dfrac{2}{Q^4-4 {m^2_t} Q^2}
\{({k_2}\cdot {{q_2}}  ) ({Q^2} {s_1}-2 {m^2_t} {s_2})\nonumber\\
&+({k_3}\cdot {{q_2}}  )({Q^2} ({s_2}-{s_1})-2 {m^2_t} {s_2})\}.
\end{align}
After the projection on primitive amplitudes, one also need to decompose $k_2$ in each form factor. Then the explicit expressions for $\mathbb{P}_H F_i$ can be written in spinor representation as
\begin{align}
\mathbb{P}_{-+} F_1 &= \left\langle k_{20}|\slashed{\varepsilon}|k_{30}\right],\nonumber\\
\mathbb{P}_{++} F_1 &= \dfrac{m_t}{\left\langle k_{2r}|k_{20}\right\rangle}\left\langle k_{2r}|\slashed{\varepsilon}|k_{30}\right],\nonumber\\
\mathbb{P}_{++} F_2 &= -\dfrac{m_t}{\left\langle k_{30}|k_{3r}\right\rangle} \left[k_{20}|\slashed{\varepsilon}|k_{3r}\right\rangle,\nonumber\\
\mathbb{P}_{-+} F_2 &= -\dfrac{m_t^2}{\left[k_{2r}|k_{20}\right]\left\langle k_{30}|k_{3r}\right\rangle} \left[k_{2r}|\slashed{\varepsilon}|k_{3r}\right\rangle,\nonumber\\
\mathbb{P}_{++} F_3 &= (k_3\cdot\varepsilon)\left[k_{20}|k_{30}\right],\nonumber\\
\mathbb{P}_{-+} F_3 &= (k_3\cdot\varepsilon)\dfrac{m_t}{\left[k_{2r}|k_{20}\right]}\left[k_{2r}|k_{30}\right],\nonumber\\
\mathbb{P}_{-+} F_4 &= -(k_3\cdot\varepsilon) \dfrac{m_t}{\left\langle k_{30}|k_{3r}\right\rangle} \left\langle k_{20}|k_{3r}\right\rangle,\nonumber\\
\mathbb{P}_{++} F_4 &= -(k_3\cdot\varepsilon) \dfrac{m_t^2}{\left\langle k_{2r}|k_{20}\right\rangle\left\langle k_{30}|k_{3r}\right\rangle}\left\langle k_{2r}|k_{3r}\right\rangle,\nonumber\\
\mathbb{P}_{+-} F_1 &= \left[k_{20}|\slashed{\varepsilon}| k_{30} \right\rangle ,\nonumber\\
\mathbb{P}_{--} F_1 &= \dfrac{m_t}{\left[ k_{2r}|k_{20}\right]} \left[k_{2r}|\slashed{\varepsilon} |k_{30}\right\rangle,\nonumber\\
\mathbb{P}_{--} F_2 &= -\dfrac{m_t}{\left[ k_{30}|k_{3r}\right]} \left\langle k_{20}|\slashed{\varepsilon}|k_{3r}\right],\nonumber\\
\mathbb{P}_{+-} F_2 &= -\dfrac{m_t^2}{\left\langle k_{2r}|k_{20}\right\rangle \left[ k_{30}|k_{3r}\right]} \left\langle k_{2r}|\slashed{\varepsilon}| k_{3r}\right],\nonumber\\
\mathbb{P}_{--} F_3 &= (k_3\cdot\varepsilon)\left\langle k_{20}|k_{30} \right\rangle,\nonumber\\
\mathbb{P}_{+-} F_3 &= (k_3\cdot\varepsilon)\dfrac{m_t}{\left\langle k_{2r}|k_{20}\right\rangle} \left\langle k_{2r}|k_{30} \right\rangle,\nonumber\\
\mathbb{P}_{+-} F_4 &= -(k_3\cdot\varepsilon) \dfrac{m_t}{\left[ k_{30}|k_{3r}\right]} \left[ k_{20}|k_{3r}\right],\nonumber\\
\mathbb{P}_{--} F_4 &= -(k_3\cdot\varepsilon) \dfrac{m_t^2}{\left[ k_{2r}|k_{20}\right]\left[ k_{30}|k_{3r}\right]} \left[ k_{2r}|k_{3r}\right].
\end{align}
For simplicity to show the coefficients $c_{i,H}$, we use $x_i$ to denote seven linear independent scalar products
\begin{align}
x_1 &= q_1 \cdot  k_2,\quad
x_2 = q_1 \cdot  k_3,\nonumber\\
x_3 &= q_2 \cdot  k_2,\quad
x_4 = q_2 \cdot  k_3,\nonumber\\
x_5 &= q_1^2,\quad
x_6 = q_2^2,\quad
x_7 = q_1 \cdot  q_2.
\end{align}
\end{multicols}
\vspace*{-0.2cm}\noindent\rule{8.35cm}{0.2pt}\rule{0.6pt}{0.4cm}
\begin{align}
c_{1,-+}  =&c_{1,++}= c_{2,--} =c_{2,+-} = \dfrac{2g^4_s {m^2_t}}{(D-2) {Q^2} \left(Q^2-4 {m^2_t}\right)}\int \mathrm{d}^D q_1 \mathrm{d}^D q_2
\dfrac{1}{
	{\mathcal D}_1
	{\mathcal D}_2
	{\mathcal D}_3
	{\mathcal D}_4
	{\mathcal D}_5
	{\mathcal D}_6
}\nonumber\\&
\times \big\{
-16 (D-2) m^2_{t} x_1^2
+8 (D-2) x_4 x_1^2
-16 (D-2) m^2_{t} x_2^2
-4 \left(D^3-16 D^2+80 D-124\right) x_1 x_4^2 \nonumber\\&
+4 \left(D^3-12 D^2+44 D-52\right) m^2_{t} x_4^2
-4 (D-3) (D-2) Q^2 \left(Q^2-4 m^2_{t}\right) x_1
-4 (D-2) Q^2 \left(Q^2-4 m^2_{t}\right) x_2 \nonumber\\&
-2 Q^2 \left(2 D^2-19 D+38\right) \left(Q^2-4 m^2_{t}\right) x_7
+2 (D-4)(D-2)
\left(Q^2-4 m^2_{t}\right) Q^2 x_5\nonumber\\&
+(D-4) \left(D^2-6 D+10\right) \left(Q^2-4 m^2_{t}\right) Q^2 x_6
+4 (D-4)
\left((D-4)^2 m^2_{t}-(D-5) Q^2\right) x_1 x_6 \nonumber\\&
+2 (D-4) \left(2 (D-4)^2 m^2_{t}-\left(D^2-10 D+26\right) Q^2\right) x_2 x_6
+2 (D-2) \left(Q^2-4 m^2_{t}\right) Q^2 \left((D-2) m^2_{t}-Q^2\right) \nonumber\\&
+ \left(4 \left(D^3-12 D^2+44 D-52\right) m^2_{t}
-2 (D-4)^2 (D-2) Q^2\right)x_3^2
+4 \left(D^3-14 D^2+64 D-92\right) x_2 x_3^2  \nonumber\\&
+16 (D-2) \left(Q^2-2 m^2_{t}\right) x_1 x_2
-8 (D-2) x_1 x_2 x_3
+8 (D-5) (D-2) x_2^2 x_3
+2 (D-4) (D-2) \left(2 (D-4) m^2_{t}-Q^2\right) x_3 x_5 \nonumber\\&
+2 (D-2) \left(2 (D-4) (D-2) m^4_{t}+\left(-4 D^2+25 D-42\right) Q^2 m^2_{t}+\left(D^2-7 D+14\right) Q^4\right) x_3 \nonumber\\&
+\left(4 (D-4) (D-2) Q^2-8 \left(2 D^2-19 D+38\right) m^2_{t}\right) x_1 x_3 \nonumber\\&
+\left(8 \left(D^3-12 D^2+51 D-70\right) m^2_{t}
-4 \left(D^3-11 D^2+43 D-58\right) Q^2\right) x_2 x_3 \nonumber\\&
+2 (D-4) (D-2) \left(2 (D-4) m^2_{t}-(D-5) Q^2\right) x_4 x_5\nonumber\\&
+2 (D-2) \left(2 (D-4) (D-2) m^4_{t}-\left(D^2-9 D+14\right) Q^2 m^2_{t}-2 Q^4\right) x_4 \nonumber\\&
+\left(4 \left(D^3-10 D^2+25 D-10\right) Q^2-8 (D-3) \left(D^2-5 D-2\right) m^2_{t}\right) x_1 x_4\nonumber\\&
+\left(8 (D-5) (D-2) Q^2
-8 \left(2 D^2-19 D+38\right) m^2_{t}\right) x_2 x_4
-8 (D-5) (D-2) x_1 x_2 x_4 \nonumber\\&
+\left(8 \left(D^3-12 D^2+44 D-52\right) m^2_{t}
-2 \left(D^3-14 D^2+56 D-72\right) Q^2\right) x_3 x_4
-4 \left(D^3-14 D^2+64 D-92\right) x_1 x_3 x_4 \nonumber\\&
+4 \left(D^3-16 D^2+80 D-124\right) x_2 x_3 x_4
+4 (D-2) \left(2 (D-4) m^2_{t}-Q^2\right) x_1 x_7
+4 (D-2)
\left(2 (D-4) m^2_{t}-(D-5) Q^2\right) x_2 x_7 \nonumber\\&
-8 (D-3) \left((D-4)^2 m^2_{t}-(D-5) Q^2\right) x_3 x_7-4 (D-3)
\left(2 (D-4)^2 m^2_{t}-\left(D^2-10 D+26\right) Q^2\right) x_4 x_7
\big\}\\ \nonumber\\
c_{2,++}  =&c_{2,-+}= c_{1,+-} =c_{1,--} =\dfrac{g^4_s}{(D-2)Q^2(Q^2-4m^2_t)}\int \mathrm{d}^D q_1 \mathrm{d}^D q_2
\dfrac{1}{
	{\mathcal D}_1
	{\mathcal D}_2
	{\mathcal D}_3
	{\mathcal D}_4
	{\mathcal D}_5
	{\mathcal D}_6
}\nonumber\\
&\times\big\{
8 (D-4)^2 (D-3) m^4_{t} x_3^2
+4 (D-2) \left(2 (D-4) (D-2) m^2_{t}+(14-3 D) Q^2\right) m^4_{t} x_4
-16 (D-2)\left(2 m^2_{t}-Q^2\right) m^2_{t}x_2^2 \nonumber\\&
-8 (D-2)^2 m^2_{t} x_3^2 x_5
-8 (D-2)^2 m^2_{t} x_4^2 x_5
-8 (D-2)^2 m^2_{t} x_1^2 x_6
-8 (D-2)^2 m^2_{t} x_2^2 x_6
-(D-4)^3 Q^2 m^2_{t} \left(4 m^2_{t}-Q^2\right) x_6 \nonumber\\&
+16 (D-2) \left(Q^2-2 m^2_{t}\right) m^2_{t}x_1^2
-2 (D-2) Q^2 m^2_{t} \left(4 m^2_{t}-Q^2\right) \left(2 (D-2) m^2_{t}-(D-4) Q^2\right)\nonumber\\&
+4 (D-4)^2
\left(2 (D-3) m^2_{t}-(D-2) Q^2\right) m^2_{t}x_4^2\nonumber\\&
+4 (D-2) \left(2 (D-4) (D-2) m^4_{t}-(D-3) (D-2)
Q^2 m^2_{t}+(D-4) Q^4\right)m^2_{t} x_3 \nonumber\\&
+4 (D-4)^2 m^2_{t} \left(4 (D-3) m^2_{t}-(D-4) Q^2\right) x_3 x_4
+16 (D-2)^2 m^2_{t} x_1 x_3 x_7
+16 (D-2)^2 m^2_{t} x_2 x_4 x_7 \nonumber\\&
-16 (D-2) \left(3 m^2_{t}-Q^2\right) x_3 x_2^2
+4 (D-2)^2
\left(Q^2-4 m^2_{t}\right) Q^2 x_7^2
-2 (D-4) (D-2) Q^2 \left(8 m^4_{t}-6 Q^2 m^2_{t}+Q^4\right) x_5\nonumber\\&
-8 (D-2)^2
\left(2 m^2_{t}-Q^2\right) x_3 x_4 x_5
+(D-6) (D-2)^2 \left(Q^2-4 m^2_{t}\right) Q^2 x_5 x_6\nonumber\\&
+2 (D-4)\left(4 (D-4)^2 m^4_{t}
+2 \left(D^2-6 D+14\right) Q^2 m^2_{t}-\left(D^2-8 D+20\right) Q^4\right) x_1 x_6\nonumber\\&
+8 (D-4) \left((D-4)^2 m^4_{t}
+(3 D-11) Q^2 m^2_{t}-(D-4) Q^4\right) x_2 x_6
-8 (D-2)^2 \left(2 m^2_{t}-Q^2\right) x_1 x_2 x_6 \nonumber\\&
+\left(4 \left(D^3-14 D^2+68 D-104\right) Q^2-8 \left(D^3-16 D^2+80 D-124\right) m^2_{t}\right) x_1 x_4^2 \nonumber\\&
+4 (D-2) Q^2
\left(4 m^2_{t}-Q^2\right) \left(2 (D-3) m^2_{t}-(D-4) Q^2\right) x_1\nonumber\\&
+ \left(8 \left(D^3-14 D^2+64 D-92\right) m^2_{t}-4 (D-4) \left(D^2-12 D+28\right) Q^2\right) x_2 x_3^2
+8 (D-2) Q^2 \left(4 m^4_{t}-5 Q^2 m^2_{t}+Q^4\right) x_2 \nonumber\\&
-16 (D-2)
\left(Q^2-2 m^2_{t}\right)^2 x_1 x_2
-16 (D-2) \left((D-1) m^2_{t}-Q^2\right) x_1 x_2 x_3 \nonumber\\&
+4 (D-4) (D-2) \left(2 (D-4) m^4_{t}
-(D-1) Q^2 m^2_{t}+Q^4\right) x_3 x_5 \nonumber\\&
+\left(-16 \left(3 D^2-21 D+38\right) m^4_{t}+16 (3 D-8) (D-4) Q^2 m^2_{t}
-8 (D-4) (D-2) Q^4\right) x_1 x_3 \nonumber\\&
+\left(16 \left(D^3-13 D^2+53 D-70\right) m^4_{t}+8 \left(5 D^2-35 D+66\right) Q^2 m^2_{t}
-8 \left(D^2-7 D+14\right) Q^4\right) x_2 x_3 \nonumber\\&
-16 (D-2) \left(Q^2-(D-1) m^2_{t}\right) x_1^2 x_4
+4 (D-4) (D-2)
\left(2 (D-4) m^4_{t}+3 Q^2 m^2_{t}-Q^4\right) x_4 x_5 \nonumber\\&
-8 \left(2 \left(D^3-7 D^2+11 D+6\right) m^4_{t}+\left(-2 D^2+23 D-54\right)
Q^2 m^2_{t}+(10-3 D) Q^4\right) x_1 x_4 \nonumber\\&
+\left(-16 \left(3 D^2-21 D+38\right) m^4_{t}+8 \left(D^2-16 D+36\right) Q^2 m^2_{t}
+16 (D-2) Q^4\right) x_2 x_4
+16 (D-2) \left(3 m^2_{t}-Q^2\right) x_1 x_2 x_4 \nonumber\\&
+\left(4 (D-4) \left(D^2-12 D+28\right) Q^2
-8 \left(D^3-14 D^2+64 D-92\right) m^2_{t}\right) x_1 x_3 x_4 \nonumber\\&
+\left(8 \left(D^3-16 D^2+80 D-124\right) m^2_{t}
-4 \left(D^3-14 D^2+68 D-104\right) Q^2\right) x_2 x_3 x_4 \nonumber\\&
+4 \left((D-6) (D-3) Q^6+(7 D-22) (5-D) m^2_{t} Q^4
+4 \left(3 D^2-21 D+38\right) m^4_{t} Q^2\right) x_7 \nonumber\\&
+8 (D-2) \left(2 (D-4) m^4_{t}-(D-1) Q^2 m^2_{t}+Q^4\right) x_1 x_7
+8 (D-2) \left(2 (D-4) m^4_{t}+3 Q^2 m^2_{t}-Q^4\right) x_2 x_7 \nonumber\\&
+2 \left(-8 (D-4)^2 (D-3) m^4_{t}
+4 \left(3 D^2-20 D+34\right) Q^2 m^2_{t}+(D-4) \left(D^2-12 D+28\right) Q^4\right) x_3 x_7 \nonumber\\&
+8 (D-2)^2 \left(2 m^2_{t}-Q^2\right)x_2 x_3 x_7 +8 (D-2)^2 \left(2 m^2_{t}-Q^2\right) x_1 x_4 x_7 \nonumber\\&
-2 \left(8 (D-4)^2 (D-3) m^4_{t}-4 \left(D^3-12 D^2+52 D-74\right) Q^2 m^2_{t}
+\left(D^3-14 D^2+68 D-104\right) Q^4\right) x_4 x_7
\big\}\\ \nonumber\\
c_{3,++}  =& c_{3,-+} = c_{4,+-} = c_{4,--} = \dfrac{4m_t g^4_s}{(D-2)(Q^3-4m^2_tQ)^2}\int \mathrm{d}^D q_1 \mathrm{d}^D q_2
\dfrac{1}{
	{\mathcal D}_1
	{\mathcal D}_2
	{\mathcal D}_3
	{\mathcal D}_4
	{\mathcal D}_5
	{\mathcal D}_6
}\nonumber\\
&\times\big\{
-2 (D-2)^2 \left(Q^2-4 m^2_t\right) Q^2 x_7^2
+4 (D-1)(D-2)^2 m^2_t x_5 x_4^2
+2 (D-2)^2
\left((D-2) Q^2-2 (D-3) m^2_t\right) x_5 x_3^2 \nonumber\\&
-2 (D-2)^2 \left(D Q^2-4 m^2_t\right) x_3 x_4 x_5
+4 (D-1) (D-2)^2
m^2_t x_6 x_2^2 +2 (D-2)^2 \left(Q^2-4 m^2_t\right) Q^2 x_5 x_6 \nonumber\\&
+2 (D-2)^2 \left((D-2) Q^2-2 (D-3) m^2_t\right) x_6 x_1^2
-2 (D-2)^2 \left(D Q^2-4 m^2_t\right) x_1 x_2 x_6
-4 (D-1)(D-2)^2 \left(2 m^2_t-Q^2\right) x_2 x_3 x_7 \nonumber\\&
+4 (D-2)^2
\left(2 (D-3) m^2_t-(D-2) Q^2\right) x_1 x_3 x_7
+4 (D-2)^2 \left(2 (D-3) m^2_t+Q^2\right) x_1 x_4 x_7
-8 (D-1)(D-2)^2 m^2_t x_2 x_4 x_7 \nonumber\\&
+2 (D-3)(D-2) \left(Q^3-4 m^2_t Q\right)^2 x_5
-2 (D-2)
\left(4 m^2_t-Q^2\right) \left(2 \left(D^2-6 D+10\right) m^2_t-(D-3) D Q^2\right) x_3 x_5\nonumber\\&
-4 (D-2)\left(4 m^2_t-Q^2\right)
\left(\left(D^2-6 D+10\right) m^2_t+(D-3) Q^2\right) x_4 x_5
+8 (D-2) m^2_t \left(4 m^2_t-Q^2\right) x_2^2 \nonumber\\&
+4 (D-2)
\left(4 m^2_t-Q^2\right) \left(2 m^2_t-(D-2) Q^2\right) x_1^2
-2 (D-4)(D-2) Q^2 \left(8 m^4_t-6 Q^2 m^2_t+Q^4\right) x_1 \nonumber\\&
+4 (D-2) \left(Q^2-4 m^2_t\right) Q^2 \left((D+4) m^2_t-2 Q^2\right) x_2
+4 (D-2)\left(4 m^2_t-Q^2\right)
\left(4 m^2_t+(D-4) Q^2\right) x_1 x_2 \nonumber\\&
-4 (D-4)(D-2)(D-1) \left(2 m^2_t-Q^2\right) x_1 x_2 x_3
+2 (D-2) m^2_t
\left(4 m^2_t-Q^2\right) \left(2 (D-2) m^2_t+(D-4)^2 Q^2\right) x_3 \nonumber\\&
+4 (D-2) \left(2 \left(D^2-5 D+12\right) m^2_t
-\left(D^2-5 D+8\right) Q^2\right) x_3 x_2^2
+4 (D-4) (D-2) (D-1)  \left(2 m^2_t-Q^2\right) x_4 x_1^2\nonumber\\&
+2 (D-2) m^2_t
\left(8 (D-2) m^4_t-2 \left(2 D^2-13 D+26\right) Q^2 m^2_t+\left(D^2-7 D+14\right) Q^4\right) x_4 \nonumber\\&
-4 (D-2)
\left(2 \left(D^2-5 D+12\right) m^2_t-\left(D^2-5 D+8\right) Q^2\right) x_1 x_2 x_4
-2 (D-2)\left(4 m^2_t-Q^2\right) \left(2 (D-4) m^2_t
+(D-3) D Q^2\right) x_1 x_7 \nonumber\\&
-2 (D-2)\left(8 (D-4) m^4_t+2 \left(-2 D^2+5 D+4\right) Q^2 m^2_t+(D-3) D Q^4\right)
x_2 x_7 \nonumber\\&
-4 \left(4 \left(5 D^2-34 D+58\right) m^2_t+\left(D^3-12 D^2+50 D-68\right) Q^2\right) x_1 x_4^2
-\left(3 D^2-20 D+36\right)
\left(4 m^2_t-Q^2\right)  m^2_t Q^2x_6\nonumber\\&
+\left(-8 (D-4)^3 m^4_t+6 \left(D^3-4 D^2+16\right) Q^2 m^2_t-\left(D^3-24 D+56\right) Q^4\right)
x_1 x_6\nonumber\\&
+\left(-8 (D-4)^3 m^4_t+2 \left(D^3-28 D^2+144 D-224\right) Q^2 m^2_t+8 \left(D^2-6 D+10\right) Q^4\right)  x_2 x_6 \nonumber\\&
+ \left(4 \left(D^3-3 D^2-8 D+28\right) m^4_t+2 (D-4)^2 (D-2) Q^2 m^2_t\right) x_3^2 \nonumber\\&
+ \left(4 (D-2) \left(D^2-6 D+10\right)
m^2_t Q^2-4 \left(D^3-9 D^2+32 D-44\right) m^4_t\right) x_4^2 \nonumber\\&
+ \left(32 (D-3) (2 D-7) m^2_t
-4 \left(D^3-3 D^2-10 D+32\right) Q^2\right) x_2x_3^2 \nonumber\\&
-4 \left(2 \left(D^3-15 D^2+76 D-116\right) m^4_t+(5 D-12) (D-6)
Q^2 m^2_t+2 (D-2) Q^4\right) x_1 x_3 \nonumber\\&
+\left(-8 (D-3) \left(D^2-24 D+60\right) m^4_t-4 \left(D^3+18 D^2-114 D+172\right)
Q^2 m^2_t+2 \left(D^3+D^2-22 D+40\right) Q^4\right) x_2 x_3 \nonumber\\&
+\left(8 \left(D^3-5 D^2-12 D+52\right) m^4_t+ 8 \left(D^3-11 D^2+47 D-70\right) Q^2 m^2_t-2 (D-4) \left(D^2-7 D+14\right) Q^4\right) x_1 x_4 \nonumber\\&
+\left(8 \left(D^3+7 D^2-68 D+116\right) m^4_t-4 \left(D^3+D^2-46 D+88\right) Q^2 m^2_t-16 (D-2) Q^4\right) x_2 x_4 \nonumber\\&
-2 \left(3 D^2-20 D+36\right) \left(D Q^2-4 m^2_t\right)m^2_t x_3 x_4
+\left(4 \left(D^3-3 D^2-10 D+32\right) Q^2
-32 (D-3) (2 D-7) m^2_t\right) x_1 x_3 x_4 \nonumber\\&
+4 \left(4 \left(5 D^2-34 D+58\right) m^2_t+\left(D^3-12 D^2+50 D-68\right) Q^2\right)
x_2 x_3 x_4 \nonumber\\&
+\left((D-6) (D-4) (D-3) Q^6+2 \left(-4 D^3+51 D^2-212 D+284\right) m^2_t Q^4
+8 \left(2 D^3-25 D^2+104 D-140\right) m^4_t Q^2\right) x_7 \nonumber\\&
+2 (D-4) \left(4 m^2_t-Q^2\right) \left(2 (D-4) (D-3) m^2_t
+(10-3 D) Q^2\right) x_3 x_7 \nonumber\\&
+2 \left(8 (D-4)^2 (D-3) m^4_t+2 \left(-3 D^3+35 D^2-140 D+184\right) Q^2 m^2_t
+\left(D^3-12 D^2+50 D-68\right) Q^4\right) x_4 x_7
\big\}\\ \nonumber\\
c_{4,-+}  =& c_{4,++} = c_{3,--} = c_{3,+-} =\dfrac{2m_tg^4_s}{(D-2)(Q^3-4m^2_tQ)^2}\int \mathrm{d}^D q_1 \mathrm{d}^D q_2
\dfrac{1}{
	{\mathcal D}_1
	{\mathcal D}_2
	{\mathcal D}_3
	{\mathcal D}_4
	{\mathcal D}_5
	{\mathcal D}_6
}\nonumber\\&\times
\big\{-4 (D-2)^2 \left(Q^2-4 m^2_t\right) Q^2 x_7^2
+8 (D-1) x_3^2 m^2_t x_5 (D-2)^2
+4  (D-2)^2
\left((D-2) Q^2-2 (D-3) m^2_t\right)x_4^2 x_5 \nonumber\\&
-4  (D-2)^2 \left(D Q^2-4 m^2_t\right) x_3 x_4 x_5 +8 (D-1) (D-2)^2 m^2_t
x_6 x_1^2 +4  (D-2)^2 \left(Q^2-4 m^2_t\right) Q^2 x_5 x_6 \nonumber\\&
+4  (D-2)^2  \left((D-2) Q^2-2 (D-3) m^2_t\right)
x_6 x_2^2-4  (D-2)^2 \left(D Q^2-4 m^2_t\right) x_1 x_2 x_6
-16 (D-1)(D-2)^2 m^2_t x_1 x_3 x_7 \nonumber\\&
+8 (D-2)^2
\left(2 (D-3) m^2_t+Q^2\right) x_2 x_3 x_7
-8 (D-1) (D-2)^2 \left(2 m^2_t-Q^2\right) x_1 x_4 x_7 \nonumber\\&
+8 (D-2)^2
\left(2 (D-3) m^2_t-(D-2) Q^2\right) x_2 x_4 x_7
-8 (D-4)(D-2) \left(2 (D+1) m^2_t-Q^2\right)  x_3 x_2^2  \nonumber\\&
-4 (D-2)\left(Q^3-4 m^2_t Q\right)^2 x_5
+16 (D-2) m^2_t \left(4 m^2_t-Q^2\right) x_1^2
+8  (D-2)\left(4 m^2_t-Q^2\right)
\left(2 m^2_t+(D-2) Q^2\right) x_2^2 \nonumber\\&
+2 (D-2)\left(Q^3-4 m^2_t Q\right)^2 \left(2 (D-2) m^2_t-(D-4) Q^2\right)
-4 (D-2)\left(Q^2-4 m^2_t\right) \left(D \left(Q^2-2 m^2_t\right)-2 Q^2\right)Q^2 x_1 \nonumber\\&
+8  (D-2)\left(4 m^2_t-Q^2\right)
\left((3 D-8) m^2_t-(D-3) Q^2\right)Q^2 x_2
-8 (D-2)\left(4 m^2_t-Q^2\right) \left(D Q^2-4 m^2_t\right) x_1 x_2 \nonumber\\&
+8 (D-2)\left(4 m^2_t-Q^2\right) \left(2 (D-3) m^2_t-Q^2\right) x_3 x_5 \nonumber\\&
-4 (D-2)\left(4 m^2_t-Q^2\right) \left(2 (D-3) (D-2) m^4_t
-(D-2) (D-1) Q^2 m^2_t+(D-4) Q^4\right) x_3 \nonumber\\&
+8 (D-2)\left(2 \left(D^2-3 D+4\right) m^2_t-D Q^2\right) x_1 x_2 x_3
+4 (D-2)\left(4 m^2_t-Q^2\right) \left(4 (D-3) m^2_t-(D-4) Q^2\right) x_4 x_5 \nonumber\\&
+8 (D-2)
\left(D Q^2-2 \left(D^2-3 D+4\right) m^2_t\right) x_4 x_1^2  \nonumber\\&
-4 (D-2)\left(4 m^2_t-Q^2\right) \left(2 (D-3) (D-2) m^4_t
+2 (D-2) Q^2 m^2_t-(D-4) Q^4\right) x_4 \nonumber\\&
+8 (D-4)(D-2) \left(2 (D+1) m^2_t-Q^2\right) x_1 x_2 x_4
-4 (D-2) \left(4 m^2_t-Q^2\right) \left(2 (D-4) m^2_t-D Q^2\right) x_1 x_7 \nonumber\\&
-4 (D-2)\left(4 m^2_t-Q^2\right)
\left(2 (D-4) m^2_t+D Q^2\right) x_2 x_7 \nonumber\\&
+\left((D-4)^2 (D-2) Q^6+2 \left(-4 D^3+43 D^2-148 D+164\right) m^2_t Q^4
+8 \left(2 D^3-23 D^2+84 D-100\right) m^4_t Q^2\right) x_6 \nonumber\\&
+\left(-16 (D-4)^3 m^4_t-4 \left(D^3-4 D^2+24 D-64\right)  Q^2 m^2_t+2 \left(D^3-8 D^2+36 D-64\right) Q^4\right) x_1 x_6 \nonumber\\&
-4 \left(4 (D-4)^3 m^4_t+\left(-5 D^3+44 D^2-168 D+240\right) Q^2 m^2_t+\left(D^3-8 D^2+30 D-44\right) Q^4\right) x_2 x_6 \nonumber\\&
+ \left(16 (D-2) \left(D^2-7 D+13\right) m^2_t Q^2
-8 \left(5 D^3-49 D^2+160 D-172\right) m^4_t\right)x_3^2 \nonumber\\&
+\left(12 (D-4)^2 (D-2) m^2_t Q^2-8 (D-3)
\left(3 D^2-28 D+52\right) m^4_t\right)x_4^2 \nonumber\\&
+ \left(32 (D-3) \left(D^2-8 D+22\right) m^2_t-16 \left(D^3-9 D^2+31 D-38\right) Q^2\right)
x_1 x_4^2 \nonumber\\&
-8 \left(4 \left(D^3-10 D^2+38 D-50\right) m^2_t+(3 D-10) (D-4) Q^2\right) x_2 x_3^2 \nonumber\\&
-4 \left(4 \left(3 D^3-23 D^2+76 D-100\right) m^4_t-4 \left(D^3-4 D^2+11 D-26\right) Q^2 m^2_t+(3 D-10) D Q^4\right) x_2 x_3 \nonumber\\&
+8 \left(2 \left(D^3-5 D^2-4 D+36\right) m^4_t+\left(-D^3+D^2+14 D-32\right) Q^2 m^2_t+(D-2)^2 Q^4\right) x_1 x_3 \nonumber\\&
+\left(16 \left(3 D^3-25 D^2+60 D-28\right) m^4_t-8 \left(D^3-12 D^2+22 D+20\right) Q^2 m^2_t-4 \left(D^2+2 D-16\right) Q^4\right)
x_1 x_4 \nonumber\\&
-8 \left(2 (D-3) \left(D^2+12\right) m^4_t+(3 D-8) (D-6) Q^2 m^2_t-(D-4) (D-2) Q^4\right) x_2 x_4 \nonumber\\&
+\left(-16 \left(4 D^3-43 D^2+148 D-164\right) m^4_t+4 \left(9 D^3-100 D^2+348 D-384\right) Q^2 m^2_t-8 (D-4)^2 (D-2) Q^4\right) x_3 x_4 \nonumber\\&
+8 \left(4 \left(D^3-10 D^2+38 D-50\right) m^2_t+(3 D-10) (D-4) Q^2\right)
x_1 x_3 x_4 \nonumber\\&
+\left(16 \left(D^3-9 D^2+31 D-38\right) Q^2-32 (D-3) \left(D^2-8 D+22\right) m^2_t\right) x_2 x_3 x_4 \nonumber\\&
-2 \left((D-6) (D-4) Q^6-2 \left(3 D^2-36 D+92\right) m^2_t Q^4+8 \left(D^2-16 D+44\right) m^4_t Q^2\right) x_7 \nonumber\\&
+4 (D-4) \left(4 m^2_t-Q^2\right) \left(2 (D-4) (D-3) m^2_t+(10-3 D) Q^2\right) x_3 x_7 \nonumber\\&
+4 \left(8 (D-4)^2 (D-3) m^4_t+2 \left(-3 D^3+35 D^2-140 D+184\right) Q^2 m^2_t+\left(D^3-12 D^2+50 D-68\right) Q^4\right) x_4 x_7\big\}.
\end{align}
\ruledown \vspace{0.5cm}
\begin{multicols}{2}
To check the above results for coefficient $c_{i,H}$, we use the Tarasov's method \cite{Tarasov:1996br} and IBP reduction to reduce the amplitude numerically. Then we can apply the numerical IBP reduction to the coefficients $c_{i,H}$ and compare these two results. During the numerical check, we choose all combinations of $(Q^2,m^2_t)\in\{(204,31),(342,76),(604,131)\}$ and $D\in\{13,17,21\}$. After the IBP reduction by LiteRed \cite{Lee:2013mka}, all numerical expressions of the coefficients $c_{i,H}$ are consistent with the numerical reduction results by Tarasov's method using FaRe \cite{Fiorentin:2015vha} and LiteRed. For reader's convenience, we show one set of explicit numerical expressions of the coefficients $c_{i,H}$ in $Q^2 = 204$, $m^2_t = 31$ and $D = 13$. Here we define
\begin{align}
I_{a_1,a_2,a_3,a_4,a_5,a_6,a_7} \equiv  \int
\dfrac{\mathrm{d}^D q_1 \mathrm{d}^D q_2}{
	{\mathcal D}^{a_1}_1
	{\mathcal D}^{a_2}_2
	{\mathcal D}^{a_3}_3
	{\mathcal D}^{a_4}_4
	{\mathcal D}^{a_5}_5
	{\mathcal D}^{a_6}_6
	{\mathcal D}^{a_7}_7
}.
\end{align}
Then the numerical coefficients are
\end{multicols}
\ruleup
\begin{align}
c_{1,-+}  =&c_{1,++}= c_{2,--} =c_{2,+-} = \nonumber\\&\dfrac{g_s^4}{547285587552000}(-349758480361050 I_{0,0,0,1,1,1,1}+12411033570133 I_{0,0,1,1,1,0,0}\nonumber\\&+561600032236650 I_{0,1,0,0,0,1,0}+54203486007512020 I_{0,1,0,1,1,1,0}-2524949213666400 I_{0,1,0,1,1,1,1}\nonumber\\&-37669046330192400 I_{0,1,1,0,0,1,0}+1725030658739520 I_{0,1,1,1,1,0,0}-4392957485551487040 I_{0,1,1,1,1,1,0}\nonumber\\&-43861242480409840 I_{0,2,0,1,1,1,0}-29063980357164172800 I_{0,2,1,1,1,1,0}+1620333018035519040 I_{1,1,0,1,1,1,1}\nonumber\\&-72885095952847200 I_{1,1,1,1,1,0,0}+11596870539170956800 I_{2,1,0,1,1,1,1}),\\\nonumber\\
c_{2,++}  =&c_{2,-+}= c_{1,+-} =c_{1,--} = \nonumber\\&\dfrac{g_s^4}{16965853214112000}(8269804372235850 I_{0,0,0,1,1,1,1}+128265951369391 I_{0,0,1,1,1,0,0}\nonumber\\&-91088086096643130 I_{0,1,0,0,0,1,0}-649848399059280020 I_{0,1,0,1,1,1,0}+1445530903043788800 I_{0,1,0,1,1,1,1}\nonumber\\&+3309765421122010800 I_{0,1,1,0,0,1,0}-48712048971181200 I_{0,1,1,1,1,0,0}+100038303369820946880 I_{0,1,1,1,1,1,0}\nonumber\\&+2255145332468200880 I_{0,2,0,1,1,1,0}+558971258163723993600 I_{0,2,1,1,1,1,0}-351050113244960902080 I_{1,1,0,1,1,1,1}\nonumber\\&+4171217340297348000 I_{1,1,1,1,1,0,0}-1701371669315481561600 I_{2,1,0,1,1,1,1}),\\\nonumber\\
c_{3,++}  =&c_{3,-+}= c_{4,+-} =c_{4,--} = \nonumber\\&\dfrac{m_tg_s^4}{169658532141120000}(15697729816313400 I_{0,0,0,1,1,1,1}-465105777241369 I_{0,0,1,1,1,0,0}\nonumber\\&+158565732883624440 I_{0,1,0,0,0,1,0}-576650937419554060 I_{0,1,0,1,1,1,0}-2324636412144847200 I_{0,1,0,1,1,1,1}\nonumber\\&-5066094845578723200 I_{0,1,1,0,0,1,0}-86512591540544400 I_{0,1,1,1,1,0,0}-14868254094936246720 I_{0,1,1,1,1,1,0}\nonumber\\&-1977030323227905200 I_{0,2,0,1,1,1,0}-159507262360225382400 I_{0,2,1,1,1,1,0}+591796539388769901120 I_{1,1,0,1,1,1,1}\nonumber\\&+544504397058468000 I_{1,1,1,1,1,0,0}+2749018573751407718400 I_{2,1,0,1,1,1,1}),\\\nonumber\\
c_{4,-+}  =&c_{4,++}= c_{3,--} =c_{3,+-} = \nonumber\\&\dfrac{m_tg_s^4}{18850948015680000}(2930558305311000 I_{0,0,0,1,1,1,1}-155537250759781 I_{0,0,1,1,1,0,0}\nonumber\\&+20515397403643560 I_{0,1,0,0,0,1,0}-89892274740750940 I_{0,1,0,1,1,1,0}-259544532715984800 I_{0,1,0,1,1,1,1}\nonumber\\&-575810327185300800 I_{0,1,1,0,0,1,0}-12234451069011600 I_{0,1,1,1,1,0,0}+16492851980585536320 I_{0,1,1,1,1,1,0}\nonumber\\&-210958059931545200 I_{0,2,0,1,1,1,0}+108826317247815014400 I_{0,2,1,1,1,1,0}+86665228257872911680 I_{1,1,0,1,1,1,1}\nonumber\\&-54851022351180000 I_{1,1,1,1,1,0,0}+401116455907129497600 I_{2,1,0,1,1,1,1}).
\end{align}
\ruledown \vspace{0.5cm}
\begin{multicols}{2}
\section{Conclusion}

In this paper, based on the massive spinor decomposition, we proposed an extended projection method to reduce the loop amplitude containing fermion chain with two massive spinors. By decomposing the massive spinor to null spinor and reference spinor, this approach can overcome the difficulty of inconsistency between helicity and chirality. To demonstrate the effectiveness of the extended projection method in high order correction, we present the tensor reduction on both one-loop and two-loop amplitudes for virtual Z boson decaying to top-quark pair. In the future, this approach can be implemented in more complicated processes including the production of multiple massive fermions.

\section{Acknowledgments}

\acknowledgments{The authors want to thank Bo Feng, Thomas Gehrmann, Gang Yang, Li Lin Yang and Hua Xing Zhu for helpful discussions.}

\end{multicols}
\vspace{-1mm}
\centerline{\rule{80mm}{0.1pt}}
\vspace{2mm}
\begin{multicols}{2}
	\bibliographystyle{cpc}
	\bibliography{massive_projection}
\end{multicols}
\clearpage
\end{document}